\documentclass[12pt]{article}
\begin{document}
\title{{Comment on ``Conjectures on exact solution of
three-dimensional (3D) simple orthorhombic Ising lattices"}%
\footnote{Supported by NSF grantPHY 07-58139}}

\author{Jacques H.H. Perk%
\footnote{Email: perk@okstate.edu}\\ \cr
145 Physical Sciences, Oklahoma State University,\\
Stillwater, OK 74078-3072, USA\footnote{Permanent address}\\
and\\
Department of Theoretical Physics, (RSPSE), and\\
Centre for Mathematics and its Applications (CMA),\\
Australian National University,\\
Canberra, ACT 2600, Australia}

\maketitle

\begin{abstract}
It is shown that a recent article by Z.-D. Zhang is in error
and violates well-known theorems.
\end{abstract}

After receiving an electronic reprint of Zhang's recent paper
\cite{zdz} some time ago, I have had an email exchange with the
author pointing out a number of errors in the paper, which
unfortunately invalidate all its main results. As now also
follow-up papers \cite{scd,km} have appeared using Zhang's
erroneous results, I felt finally compelled to write down
some of my criticism. The editor of the journal has supplied me
with copies of a competing comment \cite{wmfc} and Zhang's
reply to it. Here I shall bring up several other issues with
\cite{zdz} that are not discussed in \cite{wmfc}.

Putative ``exact solutions" of the 3d Ising model have been
advocated before, see e.g.\ \cite{mr,ddd,lw}. In 1952 J.R.\
Maddox (the later editor of Nature) showed his solution \cite{mr}
of the 3d Ising model at the StatPhys 2 conference in Paris.%
\footnote[1]{In the proceedings he is called M.\ Maddox, with
M for Monsieur (Mister in French).} The error in his calculation
was caught at the meeting and was the result of an incorrect
application of the Jordan-Wigner transformation \cite{kaufman}.
This error has been made also by Zhang in eqs.\ (15) and (16)
[(3.3) and (3.4) on pages 12 and 13]\footnote[2]{References
to equations and pages of the arXiv preprint are given within
square brackets.} of \cite{zdz}.

Using both Kaufman's original notations---see eq.\ (11) in
\cite{kaufman}---and more modern notations, the Jordan--Wigner
transformation is given by
\begin{eqnarray}
s_j\equiv&\sigma^x_j&=
\left[\prod_{k=1}^{j-1}i\Gamma_{2k-1}\Gamma_{2k}\right]
\Gamma_{2j-1},\nonumber\\
is_jC_j\equiv&\sigma^y_j&=
\left[\prod_{k=1}^{j-1}i\Gamma_{2k-1}\Gamma_{2k}\right]
\Gamma_{2j},\nonumber\\
\quad C_j\equiv&-\sigma^z_j&=i\Gamma_{2j-1}\Gamma_{2j},
\quad\Gamma_{2j-1}\equiv P_j,\quad\Gamma_{2j}\equiv Q_j.
\end{eqnarray}
It is essential that all $\Gamma$ matrices anticommute
($\Gamma_k\Gamma_l=-\Gamma_l\Gamma_k$, $k\ne l$, but
$\Gamma_k^{\;2}=1$), in order to be able to use the theory of
spinor representations of the rotation group.
In Zhang's paper \cite{zdz}, $j$ runs from 1 to $nl$,
corresponding to his $({\bf r},{\bf s})$ running from
$(1,1)$ to $(n,l)$, or  $j=(n-1)r+s$.
Then one finds,\footnote[3]{I omit here the
extra $U$-factors in the terms $j=nk$
(for $k=1,\dots,l$) resulting from periodic boundary conditions
\cite{kaufman}, which are also present in \cite{zdz}. With open
boundary conditions there are no such $U$'s.}
\begin{equation}
\sum_j \sigma^x_j\sigma^x_{j+1}=\sum_j
i\Gamma_{2j}\Gamma_{2j+1},\quad
-\sum_j \sigma^z_j=\sum_j i\Gamma_{2j-1}\Gamma_{2j},
\end{equation}
agreeing with (15a) and (15c) [(3.3a) and (3.3c)] of \cite{zdz},
but one should have
\begin{equation}
\sum_j \sigma^x_j\sigma^x_{j+n}=\sum_j i\Gamma_{2j}
\left[\prod_{k=j+1}^{j+n-1}i\Gamma_{2k-1}\Gamma_{2k}\right]
\Gamma_{2j+2n-1},
\label{err}
\end{equation}
whereas in (15b) [(3.3b)] of \cite{zdz} one finds a quadratic form
equivalent to
\begin{equation}
\sum_j \sigma^x_j\sigma^x_{j+n}=\sum_j i\Gamma_{2j}\Gamma_{2j+2n-1}.
\end{equation}
There is a corresponding error in (16) [(3.4)] of \cite{zdz}, as one
can verify that the $P$'s and the $Q$'s there do not all anticommute
as is required.\footnote[4]{Lou and Wu \cite{lw} correctly apply an
equation equivalent to (\ref{err}) (using $Z$ for $\sigma^x$ and $X$
for $-\sigma^z$). However, they go wrong after (77) in \cite{lw},
as they assume too much commutativity for the factors of $P$.
I thank Dr.\ Zhang for bringing \cite{lw} to my attention.}

This is the first major error in \cite{zdz}. Therefore,
Zhang has gotten a free fermion model in three dimensions in his
formulae (15), just like Maddox \cite{mr}. The $U$ factors in (15)
are irrelevant, since Zhang could have started with open boundary
conditions. In the thermodynamic limit for the bulk free energy
per site the boundary terms have no effect because of the
Bogoliubov inequality, as the surface to volume ratio vanishes
for the infinite system. After this error there is no need to
go to a fourth dimension as done in (17) on page 5317
[(3.5) on page 14] of \cite{zdz}; Maddox's result follows
more directly.

The high-order terms in (\ref{err}) above
and in the corresponding exponential factors of the transfer matrix
render the 3d Ising model inaccessible to exact solution.
Barry Cipra, Sorin Istrail and others have even claimed
that the 3d Ising model is NP-complete \cite{cipra}. 
This means that one should not expect a simple closed-form
solution similar to the one of the 2d Ising model in zero field
to exist.
 
One of the main results of \cite{zdz} is formula (49) for
the partition function per site on page 5325
[(3.37) on page 26], which has three
parameters given in the appendix. On page 5399 [page 137] one
finds eqs.\ (A.1), (A.2) and the following text, where these
three parameters are expressed as $w_x=1$, $w_y=w_z$ equal
to an expression with the coefficients $b_0$ through $b_{10}$
fitted such that the high-temperature series is recovered.
Therefore, this expression for the free energy contains no
more information than the known coefficients of the
high-temperature series used.

On page 5400 [page 139] Zhang insists that $w_y=w_z\equiv0$ as
soon as the temperature is finite. This is discussed further in eqs.\
(A.11)--(A.13) on pages 5405--5406 [pages 145--146], with $\kappa$
the usual high-temperature variable $\tanh K$. There is a marked
difference between the ``high-temperature limit" (A.11) and
eq.\ (A.13) for more general temperature, as the author chooses
$w_x=1$, and $w_y=w_z=0$, as soon as the temperature
is finite, which is highly inconsistent with the earlier fit.

Indeed, the procedure is clearly wrong as the convergence
of the high-temperature series has been rigorously proved
in the 1960s \cite{gms,lp} and this proof has been quoted
in many textbooks \cite{dr,rbg,sms}. This proof is based on
the proof of Gallavotti and Miracle-Sol\'e \cite{gms} of the
convergence of the fugacity expansion by a use of the
Kirkwood--Salzburg equations for the lattice-gas, which is
equivalent to the Ising model. Another theorem of Lebowitz
and Penrose \cite{lp} is then used to establish a finite
radius of convergence for the correlation functions
and the free energy expressed as series in $1/T$. They
are even real analytic up to a critical point \cite{lp,rbg}.
Paper \cite{zdz} therefore violates well-established theorems.
The statements on page 5376 [page 102] are, therefore,
manifestly wrong.

Another criticism concerns the result for the spontaneous
magnetization given in eqs.\ (102) and (103) on page 5342
[(4.28) and (4.29) on page 50].
This can be expanded as $I=1-6x^8+\ldots$, with $x=\exp(-2K)$,
with $K=J/k_{\rm B}T$. However, in Table 2 on page 5380
[page 154] one finds $I=1-2x^6+\ldots$,
taken from the well-known low-temperature series in the
literature. That $I-1$ starts with $x^8$,
corresponds to the fact that each site in 4d has 8 neighbors;
it should be $x^6$ for 6 neighbors in 3d.
Zhang's result is analytic in the low-temperature
variable $x$,
up to his critical point and it also gives the exact value
$I=1$ at $T=0$. It has a finite radius of convergence
expanded as a series in $x$. Therefore, it must agree with
the well-known series result in Table 2, which it does not.

It has also been established that in the ferromagnetic Ising
model the thermodynamic bulk limit converges to a unique state,
apart from $H=0$, $T<T_{\rm c}$ where the state can be any
convex combination of the states obtained by the infinite-volume
limits with all boundary spins up or all down, see \cite{sms}
and references quoted. One can then study an infinite hierarchy
of a discrete version of the Schwinger--Dyson equations connecting
the correlation functions with an odd number of spins in the
thermodynamic limit with all spins up on the boundary. This
way one can easily and rigorously establish the start of the
low-temperature series for the spontaneous magnetization, in
agreement with the old results in the literature. Hence,
because of the discrepancy, Zhang's result is manifestly wrong.

Much more can be said about the correlation functions,
susceptibility, and critical exponents in sections 5, 6,
and 7. Again, all the main results are in error. I
will not go into more detail as this should  already
be clear from the arguments above.

\end{document}